# Soil Characterization of Watermelon Field through Internet of Things: A New Approach to Soil Salinity Measurement

**Md. Naimur Rahman[1], Shafak Shahriar Sozol[1], Md. Samsuzzaman[1], Md. Shahin Hossin[1], Mohammad Tariqul Islam[2], S.M. Taohidul Islam[1], Md. Maniruzzaman[3]**

[1]Patuakhali Science and Technology University, Bangladesh.
[2]Center of Advanced Electronic and Communication Engineering, Universiti Kebangsaan Malaysia, Malaysia 43600
[3]ECE Discipline, Khulna University, Bangladesh.

Corresponding author: Md. Naimur Rahman (naimur.cse4th@pstu.ac.bd)

This Research was supported by the Ministry of Science and Technology, Bangladesh.

**ABSTRACT** In the modern agricultural industry, technology plays a crucial role in the advancement of cultivation. To increase crop productivity, soil require some specific characteristics. For watermelon cultivation, soil needs to be sandy and of high temperature with proper irrigation. This research aims to design and implement an intelligent IoT-based soil characterization system for the watermelon field to measure the soil characteristics. IoT based developed system measures moisture, temperature, and pH of soil using different sensors, and the sensor data is uploaded to the cloud via Arduino and Raspberry Pi, from where users can obtain the data using mobile application and webpage developed for this system. To ensure the precision of the framework, this study includes the comparison between the readings of the soil parameters by the existing field soil meters, the values obtained from the sensors integrated IoT system, and data obtained from soil science laboratory. Excessive salinity in soil affects the watermelon yield. This paper proposes a model for the measurement of soil salinity based on soil resistivity. It establishes a relationship between soil salinity and soil resistivity from the data obtained in the laboratory using artificial neural network (ANN).

## I. INTRODUCTION

About 26% of the earth's surface is land. Human lives on the terrestrial, solid Earth, comprised of bedrock and the endured bedrock called soil. Soil is a mixture of inorganic mineral particles and organic matter of contrasting size and course of action. The particles constitute roughly 50% of the soil's volume. The pores inside the soil contain water, air, and these pores are the rest of the volume [1]. Several recent studies [2]-[13] have suggested that utilizing IoT devices for real-time soil data collection and monitoring emerges as a widespread method to enhance crop productivity effectively. Bachuwar et al. have collected the data of soil moisture, light intensity, temperature, humidity from agricultural fields and monitored the data in ThingSpeak web and android applications [14]. Along with pH, temperature, and soil moisture, R. Madhumathi et al. measured N, P, K of soil as well for ensuring better yields [15]. In another study, soil information is sent to smartphones via Wi-Fi where crop suggestions are made by analyzing the values of sensors, and the crop images are monitored to see if pesticides are needed [16]. After collecting soil information via sensors, the data is sent as an SMS to the client's mobile phone via a GSM network, which is helpful in remote areas where the mobile network is poor [17]. A. Na, W. Isaac, S. Varshney et al. have used Bluetooth technology to transfer data to a nearby cellphone which is developed on the STM32 Nucleo platform [18]. The introduction of artificial intelligence specially artificial neural network has taken the field of agriculture to another level [19]-[27]. In the works of R. Singh, S. Srivastava et al., soil information like humidity, temperature, moisture, and other data responsible for plant growth is obtained and the variation of plant growth rate with respect to the intensity of light is monitored. Data is analyzed to find the suitable condition after training Logistic Regression, Linear SVC, and Gradient Boosting classifiers that achieved the accuracy of 83.33% [28]. Some case studies were used as examples to compare results, and a machine learning model was developed to make a decision on crops that would provide maximum output in a particular area. In this way, farmers can achieve higher yields and earn more profits [29]. An automated irrigation system consisting of a wireless temperature sensor and a soil moisture sensor programmed to the microcontroller has been developed to optimize water usage. ANN is used for optimizing the usage of water in this system as it predicts the water usage in the coming days. There has been more research on IoT based artificial intelligence approaches. A. Aliberti et al. present an innovative system based on Internet-of-Things (IoT) technology for forecasting the indoor air temperature of the building [30]. In detail, the methodology uses a specialized non-linear autoregressive neural network. This system makes short- and medium-term predictions, envisioning two different exploitations: (i) on realistic artificial data and (ii) on real data collected by IoT devices deployed in the building. Prediction models are trained on realistic synthetic data based on a NAR architecture with a high number of regressors. The accuracy of both real and synthetic data is investigated. This methodology has been trying to compensate the lack of real-world data in this sector. AI has been used to explore co-relationships between different soil parameters. Soil parameters like pH has been measured to determine the availability of the micronutrients in soil that suggests soil with required nutrients. This system helps to improve the yield of quality crops. There has been much research regarding the correlation between soil properties and resistivity[30]-[35]. Determining the electrical resistivity of soil will help us learn more about soil characteristics. The impact of soil properties, such as soil pH, soil salinity, and soil electrical conductivity, on the growth of cucumbers has been evaluated using the 2D Electrical Resistivity Tomography (ERT) method. A correlation between inverted electrical resistivity response, soil properties, and crop yield has been derived using regression analysis. There are many more approaches to determine soil salinity [36]-[41]. It has been observed that low resistivity could help to increase the salinity and moisture content of the soil [42]. The geophysical application of this study can enhance agricultural practices and increase crop yields. The relationship between soil electrical resistivity ratio and soil properties is crucial for obtaining reliable soil information. The soil moisture content is measured at shallow soil water content in farmland using the Continuous Vertical Electrical Sounding technique. Different types of lands were selected, and there was a heterogeneous distribution of moisture in these lands [43]. A correlation between electrical resistivity and dry density in compacted soil, which will help to measure soil compaction on site. The conventional method of estimating soil compaction is costly and tedious. The mathematical model for soil compaction measurement helps get maximum dry density and optimum moisture contents, showing fast and cost-effective performance in compacted soil monitoring. Determining the relationship between chemical properties of soil and electrical resistivity helps to learn more about soil characteristics which may improve yields [44]. The chemical properties of soil such as soil salinity, cation exchange capacity, soil pH is determined by the electrical resistivity of soil. Different tests have been performed to determine the effect of chemical characteristics through electrical resistivity, such as Self-potential (SP), four-electrode probe method, vertical electrical sounding, electrical profiling, and non-contact electromagnetic profiling [45].

Having suitable sandy soil in almost all of riverside places, Bangladesh produces a significant number of watermelons. Since the ancient days of agriculture, farmers used to see the land and decide to cultivate a crop based on their expertise and assumption that the crop could be suitable for that soil. Farmers encounter different challenges when cultivating watermelon. They become helpless when it rains unexpectedly, and even after stopping the rain, they can't sow seeds until the soil gets dried to under a certain moisture level. Sometimes they can't sow the seeds for a whole season. Also, different soil fields have distinct characteristics and need to be treated accordingly. They still have no idea about soil parameters and what values of these parameters are optimal for watermelon growth. So, the consistency in the fruit quality is hampered. The farmers use fertilizer by assumption without knowing which one is best and how much is needed. They use pesticides for disease and insect attacks and fertilizer for any other problems.

Typically, land in a certain area possesses almost similar characteristics owing to geographical factors. So, if the soil characteristics of a specific region can be measured, then farmers can decide whether the area is suitable for watermelon cultivation. pH value, temperature, moisture, and other soil factors are needed to be measured before sowing the seeds to increase the production rate on a large scale. The agriculturists determine the soil characteristics of a specific area by testing the soil in the lab and decide whether the specific area is suitable for a specific crop or not for cultivation. But these laboratory-based procedures are time consuming and costly. For measuring the soil parameters, intelligent soil characterization systems can be employed that are comprised of different wireless sensors. The system will track the data with the sensors and then send the data to the cloud, from where the farmers can obtain the most up-to-date soil characteristics via mobile application and webpage.

This research finds how suitable the land is for watermelon cultivation and what to do to make the land more preferable for cultivating watermelon using IoT based soil characterization system. This study also finds a relationship between soil salinity and soil resistivity as the sensors primarily use the electrical properties of soil to characterize the soil parameters. The change in soil salinity affects electrical characteristics like soil resistivity.

## II. SITE VISIT AND SAMPLECOLLECTION

A field visit is a form of experiential learning activity where learners are in direct contact with the field being studied. In this approach, generalization can be prepared by observation and real-life experiences. Farm field visit offer a hands-on insight into how local agriculture work. During the field visit, the research team have visited some watermelon fields at Noluabagi under Galachipa Upazila in the Patuakhali district. Almost 50 farmers harvest watermelon frequently in that village. The experiences of the field visit were beneficial and informative as the team interacted with the farmers to acquire precise facts on watermelon cultivation which is shown in Fig. 1.

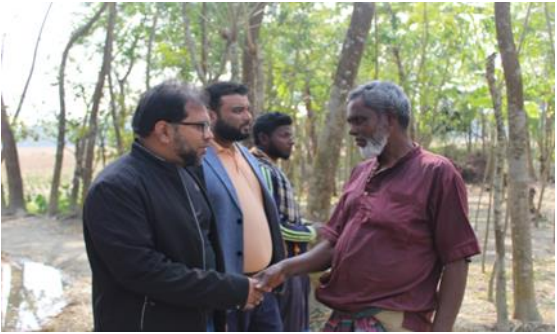

Fig. 1. Discussion with the Farmers.

A farmer cultivates around 4/5 bighas of land on average. In Bangladesh, one bigha equals to approximately 1337.8 square meters where 2500-2700 pieces of watermelons are produced in one Bigha. The research team have visited the site when the fields were ready to sow the seeds, and very tiny watermelon plants were spotted in certain fields. The team have observed the germinated seeds stored for seeding and visited the field along with the farmers, and obtained the reading of different soil characteristics using different soil meters. The research team have collaborated with the local agricultural agent to know more about watermelon cultivation and to verify the information provided by the local farmers. The team collects the watermelon cultivation information through a survey questionnaire which is shown in Table 1.

From the outcome of a successful interview with farmers about watermelon cultivation, the research team understood about site selection for preferable production, i.e., choose ideal conditional soil, land preparation, seed planting, watermelon plant insects & diseases, watermelon harvesting, etc. Mr. Hashem Ali, age 44, claims he has been cultivating watermelons for almost 20 years. He states that most farmers in this area grow "Patenga Giant" watermelon. There are several additional types as well that the farmers prefer.

Table I. Survey Questionnaire

| No. | Survey Questions (SQ) | Main Motivation |
|---|---|---|
| SQ1 | Which variety of watermelon is cultivated? | To get an idea about watermelon varieties which are cultivated most. |
| SQ2 | What kind of soil is suitable for watermelon cultivation? | To find out the suitable soil properties for watermelon cultivation. |
| SQ3 | How to prepare the soil for watermelon cultivation? | To get an idea about preparing the soil for watermelon cultivation. |
| SQ4 | Does anyone use any modern technology (sensor-based digital device) to test the suitability of soil health for watermelon cultivation? | To identify the significant modern technology used for soil testing for cultivating watermelon. |
| SQ5 | When the saline water that is washed away by the tides of the sea mixes with the soil, does cultivate watermelon in that soil harm the yield or quality of watermelon? | To investigate the effects of concentrated sea saline water soil pn watermelon cultivation. |
| SQ6 | Does anyone use any conventional method to determine soil salinity? | To know about the usual method to find out the salinity of the soil. |
| SQ7 | If this salinity harms watermelon yields or quality, does anyone take any steps to save the watermelon yield from this excess salinity? | To know about the precautionary steps taken to reduce overplus salinity. |
| SQ8 | Does anyone collect any information about cultivation using mobile communication or the internet? | To get an idea about the awareness of using information technology in agriculture. |

Sweet Baby, Top Yield, Bigtop, Champion, Ampere, World Queen, Glory, Ocean Hybrid, and Victor Super Hybrid, etc., are the most prevalent watermelon types in Bangladesh. Mr. Asir Sheikh said recently, "Black Jumbo" has been profitable for farming, and they are considering switching to this variety of watermelon to cultivate. According to him, watermelon grows especially in warm temperatures. The sandy loam type of slightly acidic soil is ideal for watermelon cultivation. Clay soil is also suitable for cultivation of watermelon. He argues that soil preparation is a crucial aspect for cultivation of watermelon. Amending the soil may increase the production of watermelon. Seaweed, aged manure, and well-rotted compost can be utilized for soil amendments.

Mr. Harun Sikdar, another watermelon farmer in the village, states that he does not know how to use the technological advantage to prepare the soil. Instead, he adopts old conventional procedures to prepare the field for watermelon cultivation. He, along with his fellow farmers, does not employ any modern technology to test the suitability of the soil for cultivation. They use their experience, and occasionally they decide on assumption. Sometimes the production is good, and sometimes it is not.

Salinity is an essential element of soil that directly influences the watermelon cultivation. Soil salinity impacts soil bacterial, fungal, and arbuscular mycorrhizal populations and subsequently affects the cultivation of watermelon [15]. The research team have discussed with farmers about the salinity issues in their fields. Mr. Imran Tarafdar said that sometimes when the weather is terrible, river water, occasionally saline water washed away by the tides of the sea, mixed with the soil and damages the quality of watermelons fields. Sometimes this increases plant mortality as well. However, they do not have any conventional procedures to measure soil salinity to overcome these limitations of cultivation. Besides, they do not have any technological support to measure the salinity of the soil. As a result, their production of watermelons is hindered. Usage of information technology among farmers is a cost-effective, efficient approach for exchanging and sharing knowledge more widely. Mr. Aroj Ali states that he owns a mobile phone, but he does not collect any information regarding cultivation from the internet or any other digital sources. Mr. Yusuf Sheikh, age 35, follows local news concerning watermelon cultivation, such as what variety of watermelon is profitable to cultivate, what the market price for watermelons is, etc.

After the interviews with the farmers, soil samples from the watermelon fields have been collected and readings of pH, moisture, and temperature using meters have been taken which are shown in Fig. 2. The team employed two types of soil-characteristics measuring meter. The Soil Temperature Meter TPJ-21 has been used to measure the temperature of soil samples. The analog lightweight Soil pH and Moisture Meter Tester (pH-707) has been used to evaluate the pH and moisture of the soil samples.

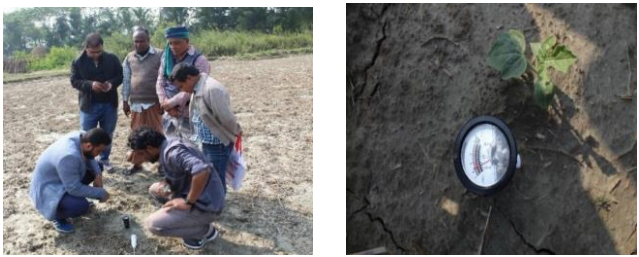

(a)                                  (b)

Fig. 2. Measuring values of soil parameters in the watermelon field.

Soil samples from the prepared watermelon field have been collected for the analysis in the laboratory. After preparing the soil for the analysis in the laboratory, the values of the parameters are to be measured, and also crosschecked with the values obtained using the developed IoT based soil parameters measuring system.

## III. METHODOLOGY

A soil characterization system has been developed to compare and analyze the field data, laboratory data and the data obtained from the IoT system. Then the necessary data is collected to establish a relationship between soil resistivity and soil salinity. The approaches adopted for this research give information on the soil physical properties of the watermelon field. The soil characterization system uses IoT technology to measure the soil properties; then the collected field soil is analyzed in the Soil Science laboratory to take data. After that a relationship between soil resistivity and soil salinity is established; and finally, an ANN model is created to show the result how strongly the soil parameters are correlated.

### A. ANALYSIS OF THE COLLECTED SOIL SAMPLES IN THE LABORATORY

The soil collected from the watermelon field is analyzed in the soil science laboratory at Patuakhali Science and Technology University. The parameters of pH, soil moisture and soil temperature are measured there. The objective of this analysis is to check the accuracy of the developed smart soil characterization system. The values acquired in the laboratory, the readings obtained by the soil characterization system, and the measurements got from the field using different meters are compared afterwards to evaluate the precision of the results. Soil characteristics identification devices like AD1000-AD1200 professional pH-ORP-TEMP Bench meter and KERRO 0-80% Digital Soil Moisture Meter have been used in the laboratory to take the temperature, moisture, and pH of the collected soil.

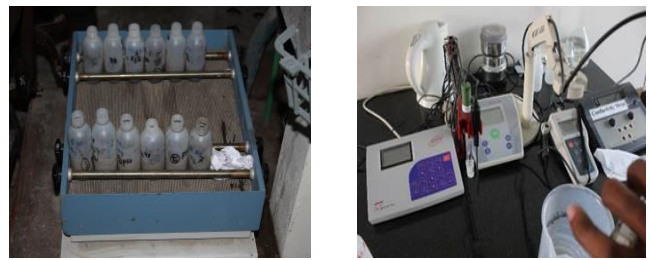

(a)                                  (b)

Fig. 3. Depicting the process of measuring soil parameters in laboratory.

As shown in Fig. 3, the soil samples collected from the field are measured by a KERRO 0-80% Digital Soil Moisture Meter to measure moisture in the laboratory. Then it is cleaned up of waste and unwanted material, and after that the soil is dried, and converted into powder. Then the soil is mixed with water. 10g of soil is taken for each sample and 20 ml of water is combined to measure the pH value. The sample mixed with water is shaken for about 45 minutes. Then the values of pH and temperature were measured and documented.

## B. IOT BASED SOIL CHARACTERIZATION SYSTEM

IoT based soil characterization system consists of different soil sensors like Waterproof DS18B20 Digital Temperature Sensor, MH-Sensor Series Flying Fish Moisture Sensor, and Soil pH Detector RS485 Analog Greenhouse High Precision 4G / NB Soil pH Sensor which are connected to the Arduino Uno. The whole circuitry is constructed on the breadboard. The power and ground of the whole circuit are taken from the raspberry pi. Then the Raspberry Pi and the Arduino Uno are connected by serial communication through the Arduino USB cable, as shown in the Fig. 4. Fig. 5 depicts the experimental work of the system to take the data of different soil parameters from the soil collected from the watermelon field.

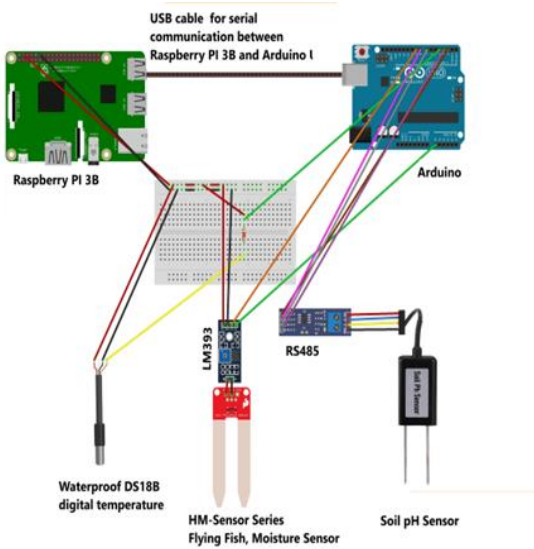

Fig. 4. Circuit Diagram of the IoT based Soil Characterization system

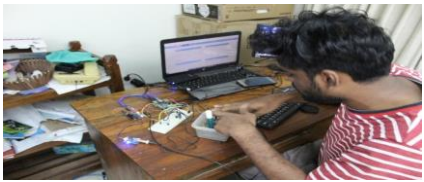

Fig. 5. Experimental work with the IoT based Soil Characterization system

The flowchart of the working of the IoT based soil characterization system is shown in Fig. 6.

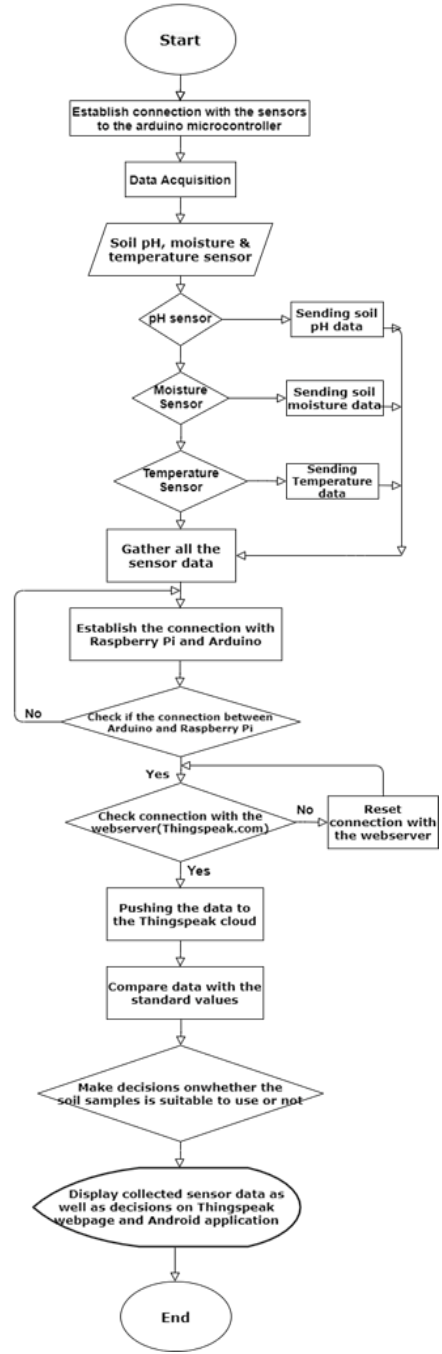

Fig. 6. Flowchart of the working of the System

ThingSpeak is an open-source Internet of Things (IoT) platform and API to store and monitor real time data online utilizing the HTTP protocol over the Internet. In the ThingSpeak channel, data can be conFig. d as public or private. In the soil characterization system, pH, moisture, and temperature of the soil sample are measured from the three analog sensors like MH-Sensor Series Flying Fish Moisture Sensor, Waterproof DS18B20 Digital Temperature Sensor, and Soil pH Detector RS485 Analog Greenhouse High Precision 4G/NB Soil pH Sensor. Analog readings from the sensors are then transformed into digital values in the Arduino Uno, which sends them to the Raspberry Pi through serial communication. The raspberry pi transmits the sensor data to the cloud server by using its inbuilt transmission module. A channel is created on the ThingSpeak server to receive data. The server provides a unique read and write key with a channel ID to communicate with the local device and to isolate the dimension of data. Once data is on a ThingSpeak channel, it could be analyzed and visualized using the MATLAB tool. The mobile application and web page, which are connected to the ThingSpeak server, then display the data. The mobile application developed for this system receives the digital data from the ThingSpeak cloud server and stores it in a local database. To acquire additional insights and depth regarding the collected data, the research team have communicated with the agriculturists and collected a reference value of pH, moisture, and temperature for a particular type of soil to determine if the sample of soil is suitable to utilize for watermelon cultivation. This data is recorded in the database of the mobile application. After collecting readings from the sensors, the mobile application analyzes the data and compares it with the standard values given by agriculturists. Then it shows if the soil sample is suitable to use or not.

## C. RELATIONSHIP BETWEEN SOIL RESISTIVITY AND SALINITY

### 1. SOIL ELECTRIC PROPERTIES

Electrical resistivity is a natural property that indicates how emphatically a given material opposes the stream of electric current. Soil electrical resistivity is viewed as a representative of the spatial and fleeting vacillation of several other soil physical parameters like structure, water content, or fluid composition. It is influenced by the physical characteristics, chemical properties, moisture content, the number of electrolytes, and ultimately, temperature. Many resistors and conductors have a uniform cross-area with a uniform progression of electric flow and are constructed of one material.

In this case, the electrical resistivity ρ (Greek: rho) is defined as:

$$\rho = \frac{RA}{L} \qquad (1)$$

Where, R is the electrical resistance of a uniform material (estimated in ohms, Ω), L is the length of the piece of material (measured in meter, m), and A is the cross-sectional area of the specimen (measured in a square meter).

Soil electric resistivity is acquired by passing the electric currents using an electric source. Electric current is conducted in the soil between the two steel probes. The amount of current that flows through the soil between the probes is straightforwardly proportionate to the electrical conductivity of the soil. The current is distinguished at the opposite end of the electrode, and reasonable voltage differences are considered to obtain resistance from soil. On the other hand, salinity is the measurement of the amount of mineral salt in the soil. Here, the salinity is measured in the unit of Percentage (%) means the quantity of salt in a particular solution or soil sample. A small bowl of known length (L) and cross-sectional area (A) and SANWA YX-360-TRF Multi Tester has been used to find the soil resistivity of the collected soil sample.

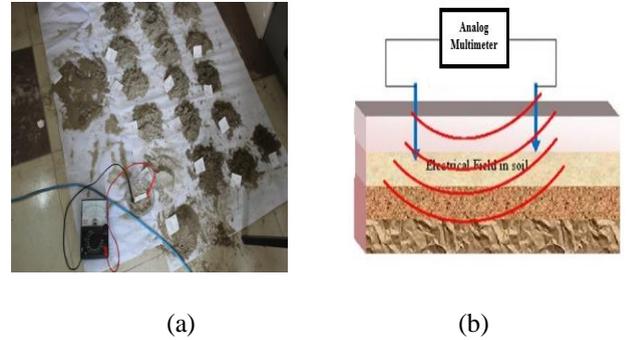

(a)                              (b)

Fig. 7: Soil resistivity measurement with analog multimeter

The generation of an electrical field in the soil provides information about the electrical conductivity and resistivity of soil. As illustrated in Fig. 7, the electrical field is generated during the passing of electrical signals from one probe to another probe through the compacted soil. The length L is measured here on a meter scale. This is the distance between the two probes of the analog multimeter. The cross-sectional area (A) of the small bowl is measured by A= $\pi r^2$ as the bowl is a round shape. The resistance (R) is obtained by the reading of the multimeter. Using these values, the resistivity is measured for several samples of the soil of different salinity.
The length between the two probes, L = 0.05 m
Radius of circular bowl = .05m.
The cross-sectional area of the bowl, $A = \pi r^2$
$$= 3.1416 \times (.05)^2$$
$$= 0.00784 \text{ m}^2$$

For 20% salinity with a fixed 5% moisture, we get the resistance (R) of these soil samples from the analog multimeter is 150 KΩ. Then we find the resistivity(ρ) from Equation (1),

$$\rho = \frac{RA}{L}$$
$$= \frac{150 \times .00784 \times 1000}{.05 \times 1000}$$
$$= 23.52 \text{ KΩm.}$$

Table I shows the measurement of Resistivity for the variation of salinity (%) of soil samples of different moisture levels.

## Table II Resistivity of soil for various salinity at different moisture level

| Water Content (%) | Soil Sample No. | Particle Size Distribution | | | Salinity (%) | Resistance, R (KΩ) | Resistivity, ρ (KΩm) |
|---|---|---|---|---|---|---|---|
| | | Sand (%) | Slit (%) | Clay (%) | | | |
| 5 | 01 | 60 | 30 | 10 | 0 | 160 | 25.088 |
| | 02 | 60 | 30 | 10 | 10 | 156 | 24.4608 |
| | 03 | 60 | 30 | 10 | 20 | 150 | 23.52 |
| | 04 | 60 | 30 | 10 | 30 | 143 | 22.4224 |
| | 05 | 60 | 30 | 10 | 40 | 130 | 20.384 |
| | 06 | 60 | 30 | 10 | 50 | 121 | 18.9728 |
| | 07 | 60 | 30 | 10 | 60 | 110 | 17.248 |
| | 08 | 60 | 30 | 10 | 70 | 104 | 16.3072 |
| | 09 | 60 | 30 | 10 | 80 | 100 | 15.68 |
| | 10 | 60 | 30 | 10 | 90 | 91 | 14.2688 |
| | 11 | 60 | 30 | 10 | 100 | 80 | 12.544 |
| 10 | 01 | 60 | 30 | 10 | 0 | 10 | 1.568 |
| | 02 | 60 | 30 | 10 | 10 | 9.7 | 1.52096 |
| | 03 | 60 | 30 | 10 | 20 | 9.5 | 1.4896 |
| | 04 | 60 | 30 | 10 | 30 | 9.2 | 1.44256 |
| | 05 | 60 | 30 | 10 | 40 | 9 | 1.4112 |
| | 06 | 60 | 30 | 10 | 50 | 8.5 | 1.3328 |
| | 07 | 60 | 30 | 10 | 60 | 8 | 1.2544 |
| | 08 | 60 | 30 | 10 | 70 | 7 | 1.0976 |
| | 09 | 60 | 30 | 10 | 80 | 6 | 0.9408 |
| | 10 | 60 | 30 | 10 | 90 | 5.5 | 0.8624 |
| | 11 | 60 | 30 | 10 | 100 | 5 | 0.784 |
| 20 | 01 | 60 | 30 | 10 | 0 | 1.2 | 0.18816 |
| | 02 | 60 | 30 | 10 | 10 | 1.1 | 0.17248 |
| | 03 | 60 | 30 | 10 | 20 | 1 | 0.1568 |
| | 04 | 60 | 30 | 10 | 30 | 0.9 | 0.14112 |
| | 05 | 60 | 30 | 10 | 40 | 0.8 | 0.12544 |
| | 06 | 60 | 30 | 10 | 50 | 0.75 | 0.1176 |
| | 07 | 60 | 30 | 10 | 60 | 0.7 | 0.10976 |
| | 08 | 60 | 30 | 10 | 70 | 0.67 | 0.105056 |
| | 09 | 60 | 30 | 10 | 80 | 0.6 | 0.09408 |
| | 10 | 60 | 30 | 10 | 90 | 0.5 | 0.0784 |
| | 11 | 60 | 30 | 10 | 100 | 0.7 | 0.10976 |
| 30 | 01 | 60 | 30 | 10 | 0 | 0.48 | 0.075264 |
| | 02 | 60 | 30 | 10 | 10 | 0.30 | 0.04704 |
| | 03 | 60 | 30 | 10 | 20 | 0.20 | 0.03136 |
| | 04 | 60 | 30 | 10 | 30 | 0.19 | 0.029792 |
| | 05 | 60 | 30 | 10 | 40 | 0.17 | 0.026656 |
| | 06 | 60 | 30 | 10 | 50 | 0.17 | 0.026656 |
| | 07 | 60 | 30 | 10 | 60 | 0.17 | 0.026656 |
| | 08 | 60 | 30 | 10 | 70 | 0.17 | 0.026656 |
| | 09 | 60 | 30 | 10 | 80 | 0.17 | 0.026656 |
| | 10 | 60 | 30 | 10 | 90 | 0.17 | 0.026656 |
| | 11 | 60 | 30 | 10 | 100 | 0.17 | 0.026656 |
| 40 | 01 | 60 | 30 | 10 | 0 | 0.20 | 0.03136 |
| | 02 | 60 | 30 | 10 | 10 | 0.16 | 0.025088 |
| | 03 | 60 | 30 | 10 | 20 | 0.14 | 0.021952 |
| | 04 | 60 | 30 | 10 | 30 | 0.14 | 0.021952 |
| | 05 | 60 | 30 | 10 | 40 | 0.14 | 0.021952 |
| | 06 | 60 | 30 | 10 | 50 | 0.14 | 0.021952 |
| | 07 | 60 | 30 | 10 | 60 | 0.14 | 0.021952 |
| | 08 | 60 | 30 | 10 | 70 | 0.14 | 0.021952 |
| | 09 | 60 | 30 | 10 | 80 | 0.14 | 0.021952 |
| | 10 | 60 | 30 | 10 | 90 | 0.14 | 0.021952 |
| | 11 | 60 | 30 | 10 | 100 | 0.14 | 0.021952 |
| 50 | 01 | 60 | 30 | 10 | 0 | 0.37 | 0.058016 |
| | 02 | 60 | 30 | 10 | 10 | 0.20 | 0.03136 |
| | 03 | 60 | 30 | 10 | 20 | 0.13 | 0.020384 |
| | 04 | 60 | 30 | 10 | 30 | 0.13 | 0.020384 |
| | 05 | 60 | 30 | 10 | 40 | 0.13 | 0.020384 |
| | 06 | 60 | 30 | 10 | 50 | 0.13 | 0.020384 |
| | 07 | 60 | 30 | 10 | 60 | 0.13 | 0.020384 |
| | 08 | 60 | 30 | 10 | 70 | 0.13 | 0.020384 |
| | 09 | 60 | 30 | 10 | 80 | 0.13 | 0.020384 |
| | 10 | 60 | 30 | 10 | 90 | 0.13 | 0.020384 |
| | 11 | 60 | 30 | 10 | 100 | 0.13 | 0.020384 |

## 2. SALINITY-RESISTIVITY RELATIONSHIP APPROXIMATION USING ANN'S

Finding a correlation between data has led to several discoveries in the history of science. But for accurate correlation, data has to be reliable and there needs to be a precise methodology. With increasing data, correlation can be identified between different characteristics that would not ordinarily be sought. The Artificial Neural Network (ANN) stands as a cutting-edge method for correlating various properties, renowned for its proficiency in mapping non-linear relationships between system inputs and outputs, while adeptly managing vast datasets [14]. Mimicking the human brain's intricate structure and functionality, ANNs excel in recognizing complex patterns and solving predictive problems. These computational models convert a set of numerical inputs into specific outputs, incorporating different configurations like single-layer or multi-layer networks and choosing between feed-forward and recurrent setups. In this research, the pH, temperature and resistivity are obtained from the soil sample and the salinity of the soil is estimated. With ANN, a correlation has been established between the gathered data from soil like pH, moisture, temperature, resistivity and salinity. In this study, the data has been acquired by using different sensors and the analog sensor readings have been transformed into digital values through a microcontroller.

The datasets focus on moisture levels of 5%, 10%, 20%, and 30%, excluding 40% and 50% due to early saturation and insufficient non-linear salinity data in resistivity measurements. The dataset has four inputs and one output:

Moisture: Soil moisture refers to the amount of water contained within the soil, crucial for plant growth, climate regulation, and ecosystem health.

pH: Soil pH is a measure of the acidity or alkalinity in soils, affecting nutrient availability, microbial activity, and plant growth.

Temperature: Soil temperature is the warmth of the soil, which influences plant growth, microbial activity, and overall ecosystem health, varying by depth, time of day, and season. It is measured in degrees Celsius.

Resistivity: Soil resistivity is a measure of the soil's ability to resist the flow of electric current, affected by moisture content, temperature, and the types of minerals present. The resistance of a soil sample is measured by a multimeter. Resistivity is obtained from multimeter readings.

Salinity: Soil salinity refers to the concentration of soluble salts in the soil, which can affect plant growth, soil structure, and microbial activity, influenced by irrigation practices, climate, and soil type.

The output of the datasets is soil salinity. It is evaluated in percentage with respect to the sample of the soil. The dataset consists of 100 samples. The dataset is split into three parts.

1.Training set: 70% of the dataset, 70 samples.
2.Validation set: 15% of the dataset, 15 samples.

3.Test set: 15% of the dataset, 15 samples.

In the described scenario, the ANN's design features three layers: an input layer, a hidden layer, and an output layer, employing a feed-forward neural network model. This model is selected for its effectiveness in accurately mapping any finite set of input-output relations. The training set of the dataset is used to train the ANN model which is used for computing the gradient and updating the weights and biases by minimizing the cost function. In training, a neural network learns to recognize the underlying association between inputs and output. The second subset is a validation set (15%). During the training process, validation error is monitored. The network weights and biases are saved at the minimum validation set error. The validation set is also employed in early halting if further training decreases the model performance. And finally, after training and validation, the test set (15%) is used to measure how well the network generalizes the data. There are different types of neural networks but, in this task, feed-forward neural network is used. In the current study, the network undergoes training using the Levenberg-Marquardt backpropagation algorithm [46]. The choice of a tan-sigmoid transfer function for the hidden layers, coupled with a linear transfer function for the output layer, is a key feature of the proposed research approach.

## IV. RESULT AND DISCUSSION

Soil characteristics data is presented and visualized on the web page and mobile application. The data collected from three approaches is compared to validate the efficiency of the system. A relationship between soil electrical resistivity and soil salinity is derived. After that, ANN is analyzed to recognize the correlation between soil salinity and soil resistivity, temperature and pH.

### A. SOIL CHARACTERIZATION RESULTS

The system sends the data to ThingSpeak server. From the server the data can be visualized by webpage. As shown in Fig. 8, data visualization of sensor readings, i. e., soil moisture, soil pH, and soil temperature in ThingSpeak web page is presented.

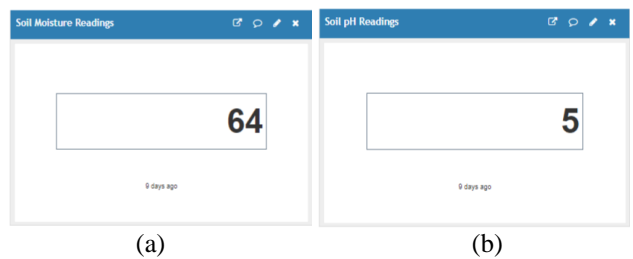

(a)                    (b)

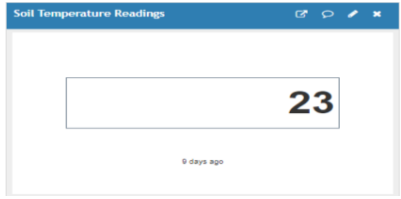

(c)

Fig. 8: Monitoring of different soil parameters on web cloud: (a) Soil moisture (b) Soil pH, (c) Soil temperature.

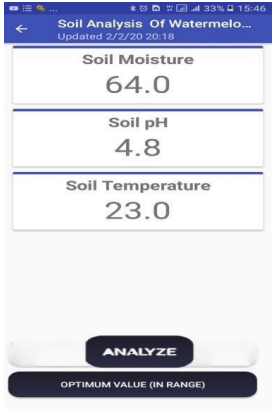

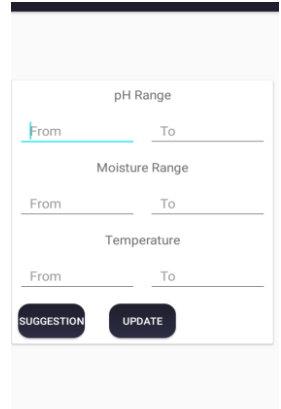

(a)                                (b)

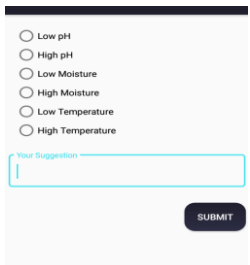

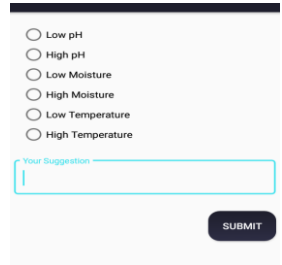

(c)                                (d)

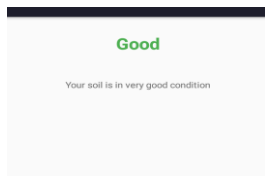

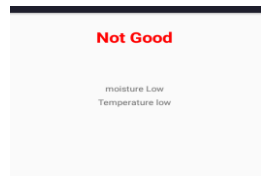

(e)                                ( f)

Fig. 9: Mobile Application Interface

Fig. 9 shows the android mobile application view and a decision architecture. From (a), it can be seen that digital sensor data from the cloud server is fetched and displayed on the application. The "OPTIMUM VALUE (IN RANGE)" button updates the standard values of pH, moisture, and temperature, which takes us to 9(b). This page is to update if there needs to be any alteration in the already set standard values. Fig. 9(c) is to give suggestions based on the analysis. The "ANALYZE" button of (a) analyzes the input values with the optimum values and outputs decision based on its analysis (d), (e). If the input values are in the range of standard values, then the soil is in good condition. Otherwise, it shows that it is bad for cultivation. It also displays the problems with the soil, i.e., "Low moisture, Low pH," etc.

*B. COMPARISON OF THE RESULTS BETWEEN THREE EXPERIMENTAL METHODOLOGIES*

The purpose of the analysis in the laboratory was to compare the values and so check the correctness and precision of characterization result. The data from the soil science laboratory, the data acquired from the field using existing meters, and the data collected through IoT based system are practically the same, as shown in Table II, III & IV. All the approaches were performed on identical soil samples from the watermelon field. The comparison of the parameters in these three approaches is shown in the following tables:

1. Soil pH comparison:

Table III: pH value comparison of the three approaches

| Sample | In the field | In laboratory | IoT system |
|--------|--------------|---------------|------------|
| A | 3.6 | 3.5 | 3.5 |
| B | 5 | 5 | 5 |
| C | 3.5 | 3.6 | 3.5 |
| D | 3.3 | 3.1 | 3.2 |
| E | 3.6 | 3.8 | 3.8 |
| F | 4 | 4.2 | 4 |

2. Soil moisture comparison:

Table IV: Soil moisture comparison of the three approaches

| Sample | In the field | In laboratory | IoT system |
|--------|--------------|---------------|------------|
| A | 70 to 80 | 75 | 80 |
| B | 60 to 70 | 64 | 64 |
| C | 80 to 90 | 86 | 85 |
| D | 80 to 90 | 87 | 86 |
| E | 80 to 90 | 85 | 85 |
| F | 60 to 70 | 65 | 65 |

3. Soil Temperature comparison:

Table V: Temperature comparison of the three approaches

| Sample | In the field | In laboratory | IoT system |
|--------|--------------|---------------|------------|
| A | 22 | 23 | 21 |
| B | 23 | 23 | 23 |
| C | 22 | 22 | 21 |
| D | 23 | 22 | 23 |
| E | 22 | 22 | 21 |
| F | 21 | 21 | 22 |

## C. RELATION BETWEEN ELECTRICAL RESISTIVITY AND SOIL SALINITY

Based on the data obtained in Table I, a relation between soil electrical resistivity and soil salinity can be established. By plotting the data and through the resistivity vs. soil salinity graph, an equation has been established.

For 5%, 10%, 20%, 30%, 40%, 50% moisture levels, the plots are drawn respectively as follows:

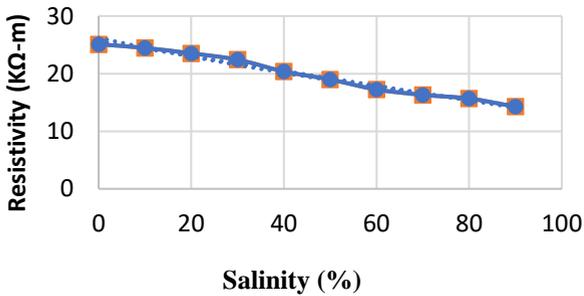

Fig. 10: Resistivity vs soil salinity for 5% moisture

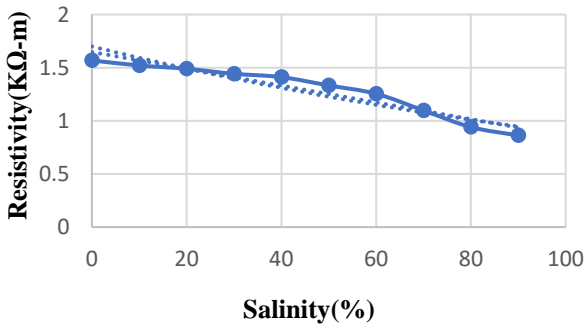

Fig. 11: Resistivity vs soil salinity for 10% moisture

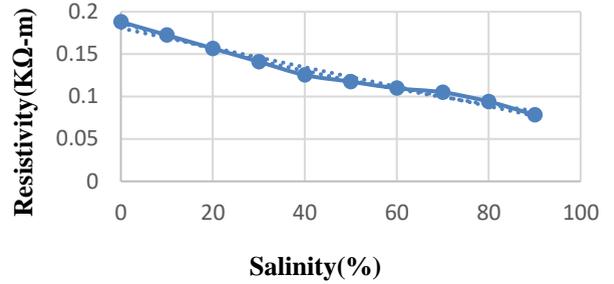

Fig. 12: Resistivity vs soil salinity for 20% moisture

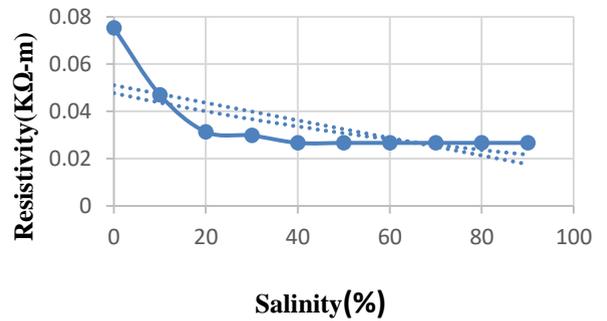

Fig. 13. Resistivity vs soil salinity for 30% moisture

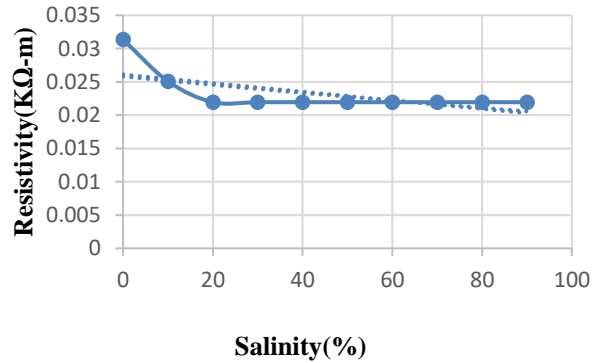

Fig. 14. Resistivity vs soil salinity for 40% moisture

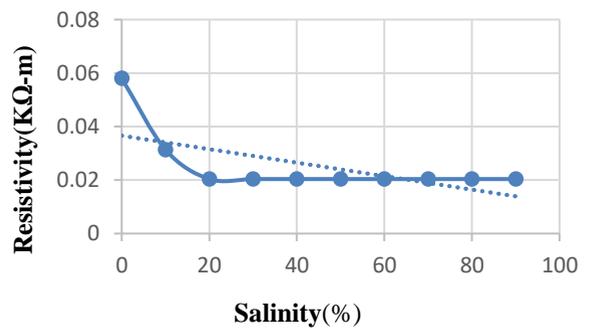

Fig. 15. Resistivity vs soil salinity for 50% moisture

The increase in salinity corresponds to decrease in resistivity and vice versa.

The equation derived for 5% moisture level is:

$$Y = 26.213e^{-0.007X} \qquad (2)$$
*Where $R^2 = 0.9802$*

The equation derived for 10% moisture level is:

$$Y = 1.7002e^{-0.007X} \qquad (3)$$
*Where $R^2 = 0.8825$*

The equation derived for 20% moisture level is:

$$Y = 0.1873e^{-0.009X} \qquad (4)$$
*Where $R^2 = 0.9913$*

The equation derived for 30% moisture level is:

$$Y = 0.0477e^{-0.009X} \qquad (5)$$
*Where $R^2 = 0.607$*

The equation derived for 40% moisture level is:

$$Y = 0.0258e^{-0.003X} \qquad (6)$$
*Where $R^2 = 0.4419$*

The equation derived for 50% moisture level is:

$$Y = 0.0332e^{-0.008X} \qquad (7)$$
*Where $R^2 = 0.4767$*

In the equations, Y represents the electrical resistivity of the soil, and X is the soil salinity. R-squared ($R^2$) is a statistical measure of how close the data is to the fitted regression line, or how well the regression predictions match the actual data points. The value of $R^2$ is closer to 1 demonstrates that the nearer the regression predictions fit the data.

To find out the equation of soil salinity from soil resistivity, equation (2), (3), (4), (5), (6), (7) can be applied with a specific moisture level respective to 5%, 10%, 20%, 30%, 40%, 50%.

For 5% moisture level from eqn. 2,

*$Y = 26.213e^{-0.007X}$*

*$R^2 = 0.9802$*

*Let, A & B an arbitrary constant.*
   *$A = 26.213$*
   *$B = 0.007$*
*Given that,*
   *$Y = Ae^{-BX}$*
 *or, $\ln Y = \ln(Ae^{-BX})$*
 *or, $\ln Y = \ln A + \ln e^{-BX}$*     *$\because \ln e = 1$*
 *or, $\ln Y - \ln A = -BX$*

*or, $BX = \ln A - \ln Y$*
*or, $BX = \ln \frac{A}{Y}$*
*or, $X = \frac{1}{B} \ln \frac{A}{Y}$*
*$\therefore X = \frac{1}{0.007} \ln \frac{26.213}{Y}$*    *(8)*

Subsequently, for 10% moisture level from eqn. 3,

$$\therefore X = \frac{1}{0.006} \ln \frac{1.6843}{Y} \qquad (9)$$

For 20% moisture level from eqn. 4,

$$\therefore X = \frac{1}{0.009} \ln \frac{0.1873}{Y} \qquad (10)$$

For 30% moisture level from eqn. 5,

$$\therefore X = \frac{1}{0.009} \ln \frac{0.0477}{Y} \qquad (11)$$

For 40% moisture level from eqn. 6,

$$\therefore X = \frac{1}{0.003} \ln \frac{0.0258}{Y} \qquad (12)$$

For 50% moisture level from eqn. 7,

$$\therefore X = \frac{1}{0.008} \ln \frac{0.0332}{Y} \qquad (13)$$

These findings investigate the relationship between soil resistivity and soil salinity across different moisture levels, revealing a significant variation in this relationship as moisture content increases. It's well-documented that soil resistivity decreases with an increase in moisture, which naturally impacts the mathematical models representing these relationships. As a result, the equations defining the relationship between soil resistivity and salinity vary across different moisture levels, reflecting the influence of moisture on the soil's electrical properties.

Many soil sensors operate based on soil electrical properties; establishing a strong relationship between resistivity (an electrical property) and soil salinity is critical. Our research has successfully formulated equations for each moisture level, ranging from 5% to 50%, each showcasing a unique exponential relationship between soil resistivity and salinity. These models provide a quantitative method for assessing soil salinity based on the soil's electrical resistivity, enabling a more precise evaluation of soil salinity based on resistivity measurements.

## C. ANN correlation output

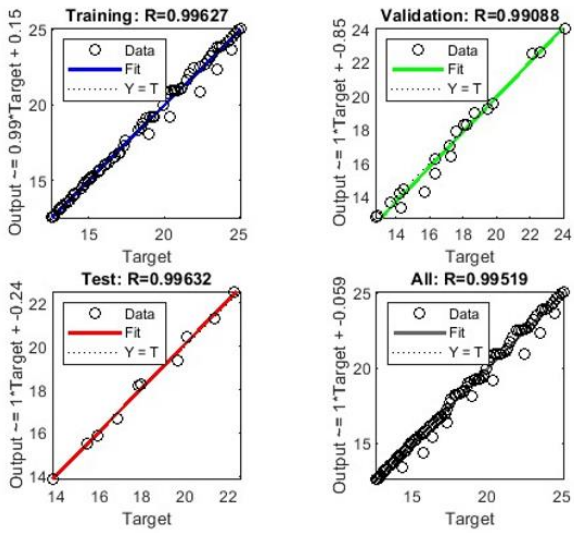

(a) 5% moisture level

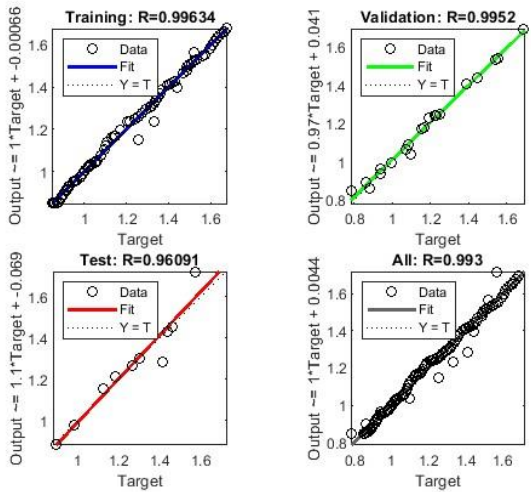

(b) 10% moisture level

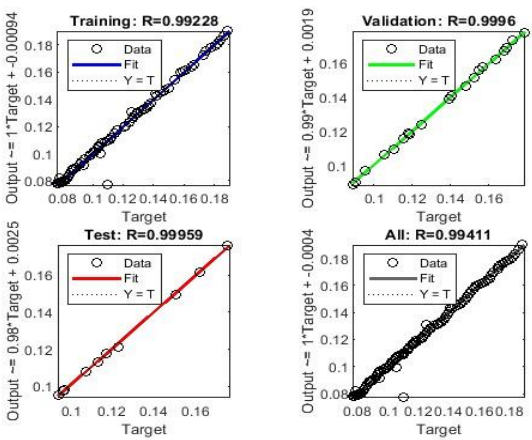

(c) 20% moisture level

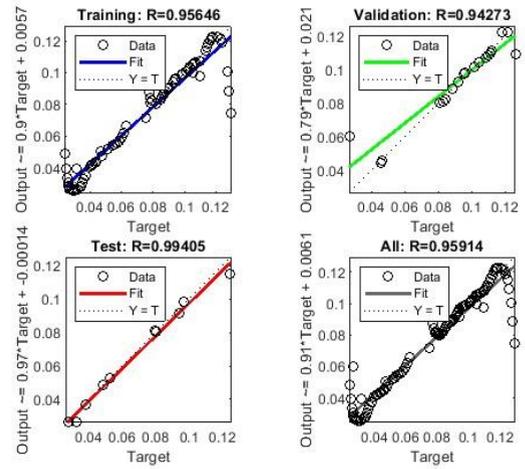

(d) 30% moisture level

Fig. 16. Regression plots for different prediction of soil salinity of definite moisture level (a) 5%, (b) 10%, (c) 20%, (d) 30%

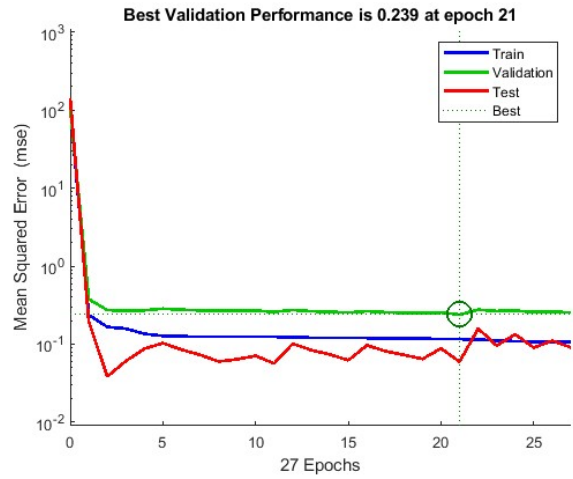

(a) MSE vs Epoch at 5% moisture level.

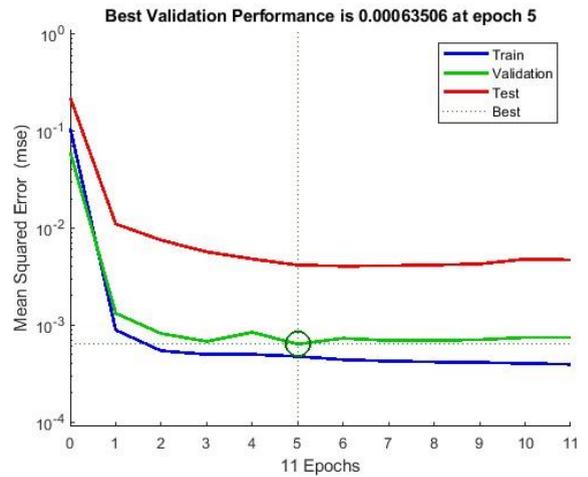

(b) MSE vs Epoch at 10% moisture level.

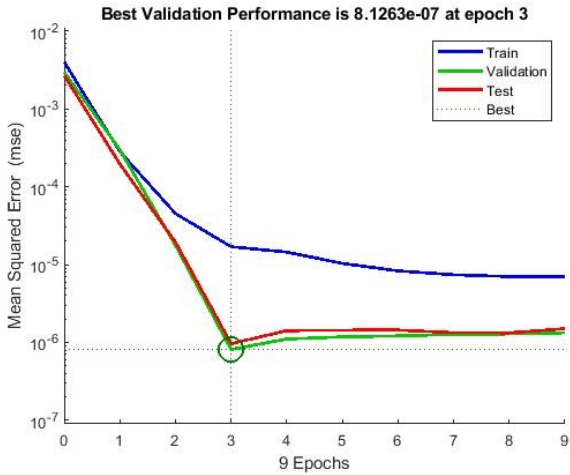

(c) MSE vs Epoch for 20% moisture level.

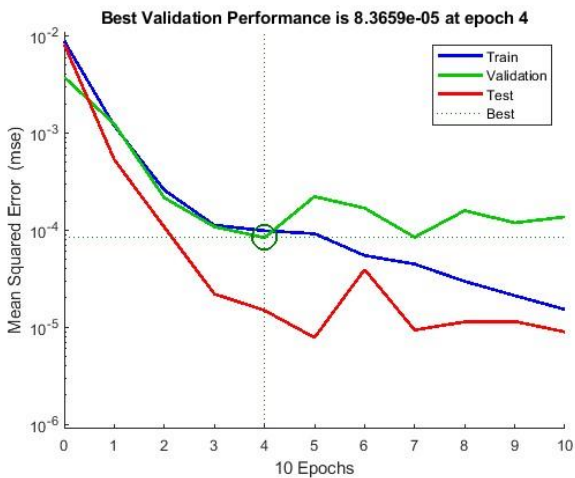

(d) MSE vs Epoch at 30% moisture level.

Fig. 17. Best validation performance. (a) MSE vs Epoch for 5% moisture level. (b) MSE vs Epoch for 10% moisture level. (c) MSE vs Epoch for 20% moisture level. (d) MSE vs Epoch for 30% moisture level.

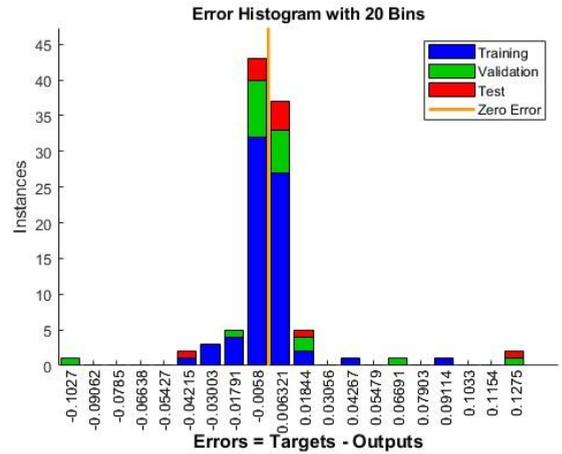

(b) Error histogram at 10% moisture

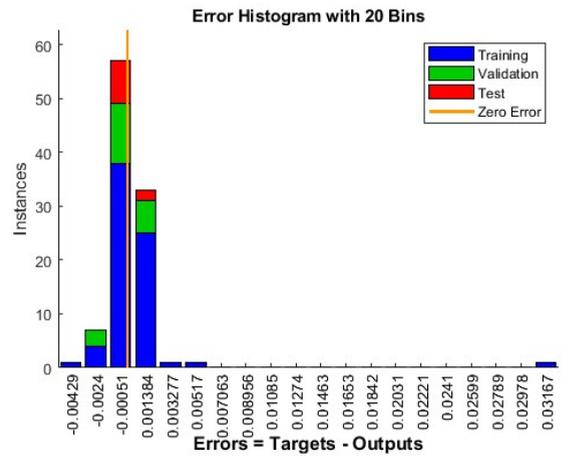

c) Error histogram at 20% moisture

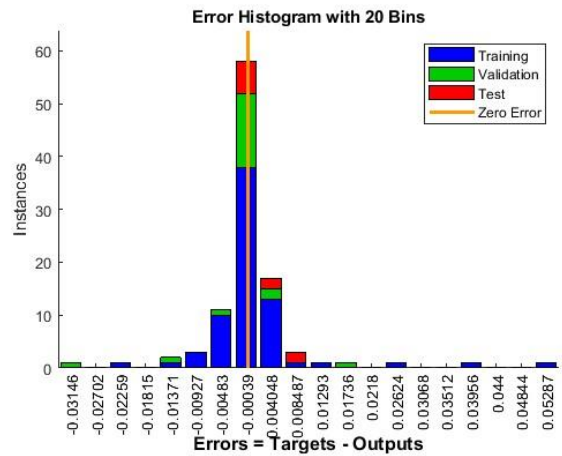

d) Error histogram at 30% moisture

Fig. 18: Characteristics of ANN plots: (a) Error histogram at 5% moisture, (b) Error histogram at 10% moisture, c) Error histogram at 20% moisture, d) Error histogram at 30% moisture

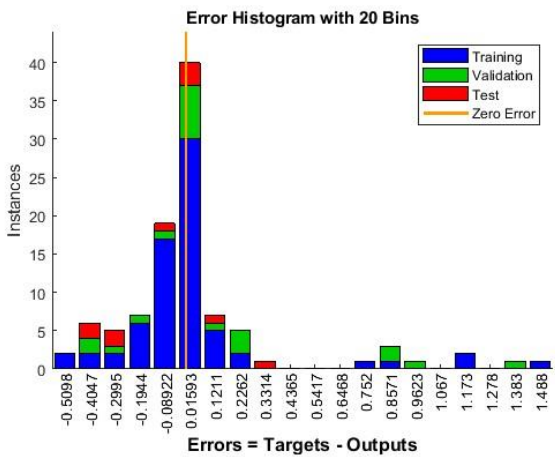

(a) Error histogram at 5% moisture

In the MATLAB environment, nntraintool is used to train the ANN on dataset. The comprehensive analysis of soil salinity using Artificial Neural Networks (ANNs) is demonstrated through regression plots (Fig. 16), learning performance over epochs (Fig. 17), and error histograms (Fig. 18), each offering critical insights into the model's predictive capabilities.

Regression plots reveal the ANN's exceptional fit to the training data across varying moisture levels, as indicated by high $R^2$ values. At 5% moisture (Fig. 16(a)), the model yields an $R^2$ of 0.99627 for training, 0.99088 for validation, 0.99632 for testing, and an overall of 0.99519, highlighting its precision in estimating soil salinity with the given inputs. The trend of high accuracy persists at 10% moisture (Fig. 16(b)) with $R^2$ values of 0.99634 for training, 0.9952 for validation, and 0.993 overall, although a slight dip in the test $R^2$ to 0.96091 suggests potential room for optimization. The 20% moisture (Fig. 16(c)) level maintains near-perfect $R^2$ Fig. s, with 0.99228 for training, 0.9996 for validation, and 0.99959 for testing, culminating in an overall $R^2$ of 0.99411. At 30% moisture (Fig. 16(d)), a marginal fluctuation in $R^2$ values is observed, with 0.95646 for training, 0.94273 for validation, and a rebound to 0.99405 in testing, with an overall value of 0.95914. These values collectively underscore the ANN's robustness and underline the input variables' significance in the salinity prediction process.

The learning performance, captured through the mean squared error (MSE) against epochs, displays the ANN's learning efficiency and adjustment to the data across moisture levels. For a 5% moisture level (Fig. 17(a)), the best validation performance was at an MSE of 0.239 at epoch 21, signifying a competent learning rate. An impressively low best validation MSE of 0.00063506 at epoch 5 for the 10% moisture level (Fig. 17(b)) indicates an exceptionally good fit of the model from an early stage. The 20% moisture level (Fig. 17(c)) showed the best validation performance with an MSE of 8.1263e-07 at epoch 3, reflecting quick learning and pattern recognition by the ANN. At the 30% moisture level (Fig. 17(d)), the model quickly adapts again, with the best validation MSE recorded at 8.3659e-05 at epoch 4. The differences in the number of epochs to achieve best validation performance at each moisture level highlight the ANN's adaptability and point to varying complexities within the data.

The error histograms provide a visual representation of the predictive accuracy at different moisture levels, showing a concentration of predictions around the zero-error mark. For the 5% moisture level (Fig. 18(a)), about 35 instances hit the zero-error bin, with errors predominantly between -0.1 and +0.1. The 10% moisture level (Fig. 18(b)), sees over 40 instances in the zero-error category, with a tighter error spread, suggesting enhanced accuracy. The 20% moisture level (Fig. 18(c)), depicts around 50 instances near zero error, with a slight broadening in the error range to -0.02 to +0.02. The 30% moisture level (Fig. 18(d)), also shows around 50 instances at zero error but with a wider distribution, indicating an increase in prediction variability as the moisture content rises.

This ANN exhibits a potent capacity for accurate soil salinity prediction, with its performance reflected in the high $R^2$ values, low MSEs, and concentrated error distributions around zero. Nevertheless, the subtle increase in error spread at higher moisture contents suggests a need for further model refinement.

## V. CONCLUSION

This research has developed and implemented a soil characterization system for watermelon cultivation. The IoT system measures the values of soil parameters that are significant for watermelon cultivation. These values are taken from time to time and transmitted to the ThingSpeak server. The real-time values are shown on a webpage. An android mobile application is also developed to make the system more user-friendly and efficient. After collecting the real-time data from the server, the mobile application gives a decision based on previously specified standard values of the parameters by an expert agriculturist. The decision is given whether the soil is good or bad for watermelon production. Based on the decision, some suggestions are also displayed on the application. The readings from the system are compared with laboratory analyzed readings of the same soil to validate the IoT system's accuracy. Also, during comparison, readings acquired by some field meters and sensors are used. In the comparison, the results achieved by each strategy are almost identical. Most soil sensors leverage the electrical characteristics of soil. So, a relationship between soil resistivity and soil salinity has been established using mathematical equations as well as ANN models. Utilizing this relationship, soil salinity has been determined by the electrical resistivity of the soil, which would help to determine the quality of the soil. This effective and resilient approach to soil characterization is beneficial to increasing production, minimizing the wastage of natural resources, and reducing the hassles for farmers.

# REFERENCES

[1] A. Badhe, S. Kharadkar, R. Ware, and P. Kamble, "IOT Based Smart Agriculture And Soil Nutrient Detection System," pp. 774–777, 2018.

[2] S. Shreya, K. Chatterjee, and A. Singh, "BFSF: A secure IoT based framework for smart farming using blockchain," *Sustainable Computing: Informatics and Systems*, vol. 40, Dec. 2023, doi: 10.1016/j.suscom.2023.100917.

[3] A. Morchid, I. G. Muhammad Alblushi, H. M. Khalid, R. El Alami, S. R. Sitaramanan, and S. M. Muyeen, "High-technology agriculture system to enhance food security: A concept of smart irrigation system using Internet of Things and cloud computing," *Journal of the Saudi Society of Agricultural Sciences*, 2024, doi: 10.1016/j.jssas.2024.02.001.

[4] H. Shahab, M. Iqbal, A. Sohaib, F. Ullah Khan, and M. Waqas, "IoT-based agriculture management techniques for sustainable farming: A comprehensive review," *Comput Electron Agric*, vol. 220, p. 108851, May 2024, doi: 10.1016/j.compag.2024.108851.

[5] J. Xu, B. Gu, and G. Tian, "Review of agricultural IoT technology," *Artificial Intelligence in Agriculture*, vol. 6. KeAi Communications Co., pp. 10–22, Jan. 01, 2022. doi: 10.1016/j.aiia.2022.01.001.

[6] S. Vijh, Arpita, J. P. Bora, P. K. Gupta, and S. Kumar, "IOT Based Real-Time Monitoring System for Precision Agriculture," in *2024 14th International Conference on Cloud Computing, Data Science & Engineering (Confluence)*, IEEE, Jan. 2024, pp. 53–58. doi: 10.1109/Confluence60223.2024.10463399.

[7] N. Chamara, M. D. Islam, G. (Frank) Bai, Y. Shi, and Y. Ge, "Ag-IoT for crop and environment monitoring: Past, present, and future," *Agricultural Systems*, vol. 203. Elsevier Ltd, Dec. 01, 2022. doi: 10.1016/j.agsy.2022.103497.

[8] S. Rudrakar and P. Rughani, "IoT based Agriculture (Ag-IoT): A detailed study on Architecture, Security and Forensics," *Information Processing in Agriculture*. China Agricultural University, 2023. doi: 10.1016/j.inpa.2023.09.002.

[9] R. Aarthi, D. Sivakumar, and V. Mariappan, "Smart Soil Property Analysis Using IoT: A Case Study Implementation in Backyard Gardening," in *Procedia Computer Science*, Elsevier B.V., 2022, pp. 2842–2851. doi: 10.1016/j.procs.2023.01.255.

[10] D. B. Anil Kumar, N. Doddabasappa, B. Bairwa, C. S. Anil Kumar, G. Raju, and Madhu, "IoT-based Water Harvesting, Moisture Monitoring, and Crop Monitoring System for Precision Agriculture," in *2nd IEEE International Conference on Distributed Computing and Electrical Circuits and Electronics, ICDCECE 2023*, Institute of Electrical and Electronics Engineers Inc., 2023. doi: 10.1109/ICDCECE57866.2023.10150893.

[11] R. Thirisha *et al.*, "Precision Agriculture: IoT Based System for Real-Time Monitoring of Paddy Growth," in *2023 International Conference on Sustainable Emerging Innovations in Engineering and Technology, ICSEIET 2023*, Institute of Electrical and Electronics Engineers Inc., 2023, pp. 247–251. doi: 10.1109/ICSEIET58677.2023.10303483.

[12] K. Sakthi, Y. M. S. Zain, S. M. S. Raj, and M. Manickavasakar, "IoT based Soil Monitoring and Control Systems," in *Proceedings - 5th International Conference on Smart Systems and Inventive Technology, ICSSIT 2023*, Institute of Electrical and Electronics Engineers Inc., 2023, pp. 516–522. doi: 10.1109/ICSSIT55814.2023.10061141.

[13] S. R. Laha, B. K. Pattanayak, S. Pattnaik, D. Mishra, D. S. Kumar Nayak, and B. B. Dash, "An IOT-Based Soil Moisture Management System for Precision Agriculture: Real-Time Monitoring and Automated Irrigation Control," in *Proceedings of the 4th International Conference on Smart Electronics and Communication, ICOSEC 2023*, Institute of Electrical and Electronics Engineers Inc., 2023, pp. 451–455. doi: 10.1109/ICOSEC58147.2023.10276266.

[14] V. D. Bachuwar, A. D. Shligram, and L. P. Deshmukh, "Monitoring the soil parameters using IoT and Android based application for smart agriculture," *AIP Conf. Proc.*, vol. 1989, 2018, doi: 10.1063/1.5047679.

[15] R. Madhumathi, T. Arumuganathan, and R. Shruthi, "Soil NPK and Moisture analysis using Wireless Sensor Networks," *2020 11th Int. Conf. Comput. Commun. Netw. Technol. ICCCNT 2020*, 2020, doi: 10.1109/ICCCNT49239.2020.9225547.

[16] N. Ananthi, J. Divya, M. Divya, and V. Janani, "IoT based smart soil monitoring system for agricultural production," *Proc. - 2017 IEEE Technol. Innov. ICT Agric. Rural Dev. TIAR 2017*, vol. 2018-Janua, pp. 209–214, 2018, doi: 10.1109/TIAR.2017.8273717.

[17] S. Biswas, "A remotely operated Soil Monitoring System : An Internet of Things ( IoT ) Application," *Int. J. Internet Things Web Serv.*, vol. 3, pp. 32–38, 2018.

[18] A. Na, W. Isaac, S. Varshney, and E. Khan, "An IoT based system for remote monitoring of soil characteristics," *2016 Int. Conf. Inf. Technol. InCITe 2016 - Next Gener. IT Summit Theme - Internet Things Connect your Worlds*, no. October, pp. 316–320, 2017, doi: 10.1109/INCITE.2016.7857638.

[19] E. Grossi and M. Buscema, "Introduction to artificial neural networks," *European Journal of Gastroenterology and Hepatology*, vol. 19, no. 12. pp. 1046–1054, Dec. 2007. doi: 10.1097/MEG.0b013e3282f198a0.

[20] P. N. Rekha, R. Gangadharan, S. M. Pillai, G. Ramanathan, and A. Panigrahi, "Hyperspectral image processing to detect the soil salinity in coastal watershed," in *4th International Conference on Advanced Computing, ICoAC 2012*, 2012. doi: 10.1109/ICoAC.2012.6416859.

[21] F. I. Siddiqui, D. M. Pathan, S. B. A. B. S. Osman, M. A. Pinjaro, and S. Memon, "Comparison between regression and ANN models for relationship of soil properties and electrical resistivity," *Arabian Journal of Geosciences*, vol. 8, no. 8, pp. 6145–6155, Aug. 2015, doi:

10.1007/s12517-014-1637-y.

[22] H. Merdun, Ö. Çinar, R. Meral, and M. Apan, "Comparison of artificial neural network and regression pedotransfer functions for prediction of soil water retention and saturated hydraulic conductivity," *Soil Tillage Res*, vol. 90, no. 1–2, pp. 108–116, Nov. 2006, doi: 10.1016/j.still.2005.08.011.

[23] M. Ebrahimi, A. A. Safari Sinegani, M. R. Sarikhani, and S. A. Mohammadi, "Comparison of artificial neural network and multivariate regression models for prediction of Azotobacteria population in soil under different land uses," *Comput Electron Agric*, vol. 140, pp. 409–421, Aug. 2017, doi: 10.1016/j.compag.2017.06.019.

[24] J. Lai, J. Qiu, Z. Feng, J. Chen, and H. Fan, "Prediction of Soil Deformation in Tunnelling Using Artificial Neural Networks," *Computational Intelligence and Neuroscience*, vol. 2016. Hindawi Limited, 2016. doi: 10.1155/2016/6708183.

[25] E. Arel, "Predicting the spatial distribution of soil profile in Adapazari/Turkey by artificial neural networks using CPT data," *Comput Geosci*, vol. 43, pp. 90–100, Jun. 2012, doi: 10.1016/j.cageo.2012.01.021.

[26] A. Bardhan and P. G. Asteris, "Application of hybrid ANN paradigms built with nature inspired meta-heuristics for modelling soil compaction parameters," *Transportation Geotechnics*, vol. 41, Jul. 2023, doi: 10.1016/j.trgeo.2023.100995.

[27] R. Al-saffar and S. Khattab, "Prediction of soil's compaction parameter using artificial neural network," *Al-Rafidain Engineering Journal (AREJ)*, vol. 21, no. 3, pp. 15–27, 2013.

[18] R. Singh, S. Srivastava, and R. Mishra, "Increasing the Yield in Crop Production," pp. 301–305, 2020.

[29] C. Dewi and R. C. Chen, *Decision making based on IoT data collection for precision agriculture*, vol. 830. Springer International Publishing, 2020.

[30] C. Kishor, H. U. Sunil Kumar, H. S. Praveena, S. P. Kavya, G. B. Apeksha, and D. J. Nayaka, "Water usage approximation of Automated Irrigation System using IOT and ANN's," *Proc. Int. Conf. I-SMAC (IoT Soc. Mobile, Anal. Cloud), I-SMAC 2018*, pp. 76–80, 2019, doi: 10.1109/I-SMAC.2018.8653692.

[31] N. K. Kishore and M. Bhagat, "Study of soil resistivity variation with salinity," in *First International Conference on Industrial and Information Systems*, IEEE, 2006, pp. 1–5.

[32] C. Y. Cordero-Vázquez, O. Delgado-Rodríguez, R. Cisneros-Almazán, and H. J. Peinado-Guevara, "Determination of Soil Physical Properties and Pre-Sowing Irrigation Depth from Electrical Resistivity, Moisture, and Salinity Measurements," *Land (Basel)*, vol. 12, no. 4, Apr. 2023, doi: 10.3390/land12040877.

[33] E. Leger, B. Dafflon, F. Soom, J. Peterson, C. Ulrich, and S. Hubbard, "Quantification of Arctic Soil and Permafrost Properties Using Ground-Penetrating Radar and Electrical

Resistivity Tomography Datasets," *IEEE J Sel Top Appl Earth Obs Remote Sens*, vol. 10, no. 10, pp. 4348–4359, Oct. 2017, doi: 10.1109/JSTARS.2017.2694447.

[34] A. Samouëlian, I. Cousin, A. Tabbagh, A. Bruand, and G. Richard, "Electrical resistivity survey in soil science: A review," *Soil and Tillage Research*, vol. 83, no. 2. pp. 173–193, Sep. 2005. doi: 10.1016/j.still.2004.10.004.

[35] V. K. Bodasingi, B. Rao, and H. K. Pillai, "Low-cost Soil Moisture and EC Sensor Design for Soil Salinity Assessment," in *2023 19th IEEE International Colloquium on Signal Processing and Its Applications, CSPA 2023 - Conference Proceedings*, Institute of Electrical and Electronics Engineers Inc., 2023, pp. 162–167. doi: 10.1109/CSPA57446.2023.10087387.

[36] A. Das, B. K. Bhattacharya, R. Setia, G. Jayasree, and B. Sankar Das, "A novel method for detecting soil salinity using AVIRIS-NG imaging spectroscopy and ensemble machine learning," *ISPRS Journal of Photogrammetry and Remote Sensing*, vol. 200, pp. 191–212, Jun. 2023, doi: 10.1016/j.isprsjprs.2023.04.018.

[37] R. Ghasempour, M. T. Aalami, S. M. Saghebian, and V. S. O. Kirca, "Analysis of spatiotemporal variations of drought and soil salinity via integrated multiscale and remote sensing-based techniques (Case study: Urmia Lake basin)," *Ecol Inform*, vol. 81, Jul. 2024, doi: 10.1016/j.ecoinf.2024.102560.

[38] S. K. Sarkar, R. R. Rudra, M. S. Nur, and P. C. Das, "Partial least-squares regression for soil salinity mapping in Bangladesh," *Ecol Indic*, vol. 154, Oct. 2023, doi: 10.1016/j.ecolind.2023.110825.

[39] A. Rafik *et al.*, "Soil Salinity Detection and Mapping in an Environment under Water Stress between 1984 and 2018 (Case of the Largest Oasis in Africa-Morocco)," *Remote Sens (Basel)*, vol. 14, no. 7, Apr. 2022, doi: 10.3390/rs14071606.

[40] F. B. T. Silatsa and F. Kebede, "A quarter century experience in soil salinity mapping and its contribution to sustainable soil management and food security in Morocco," *Geoderma Regional*, vol. 34. Elsevier B.V., Sep. 01, 2023. doi: 10.1016/j.geodrs.2023.e00695.

[41] C. Kobayashi *et al.*, "Estimating soil salinity using hyperspectral data in the Western Australian wheat belt," in *International Geoscience and Remote Sensing Symposium (IGARSS)*, 2013, pp. 4325–4328. doi: 10.1109/IGARSS.2013.6723791.

[42] A. Acheampong Aning and K. Sackey, "Implications Of Soil Resistivity Measurements Using The Electrical Resistivity Method: A Case Study Of A Maize Farm U.... Implications Of Soil Resistivity Measurements Using The Electrical Resistivity Method: A Case Study Of A Maize Farm Under Different S," *Int. J. Sci. Technol. Res.*, vol. 4, no. January, p. 12, 2015, [Online]. Available: www.ijstr.org.

[43] T. Islam, Z. Chik, M. M. Mustafa, and H. Sanusi, "Modeling of electrical resistivity and maximum dry density in soil compaction measurement," *Environ. Earth

*Sci.*, vol. 67, no. 5, pp. 1299–1305, 2012, doi: 10.1007/s12665-012-1573-7.

[44] Z. Wen-wen, W. Chong, X. U. E. Rui, and W. Li-jie, "Effects of salinity on the soil microbial community and soil fertility," *J. Integr. Agric.*, vol. 18, no. 6, pp. 1360–1368, 2019, doi: 10.1016/S2095-3119(18)62077-5.

[45] Z. Chik and T. Islam, "Study of chemical effects on soil compaction characterizations through electrical conductivity," *Int. J. Electrochem. Sci.*, vol. 6, no. 12, pp. 6733–6740, 2011.

[46] B. Yang, Y. Chen, Z. Guo, J. Wang, C. Zeng, D. Li, H. Shu, J. Shan, T. Fu, and X. Zhang, "Levenberg-Marquardt backpropagation algorithm for parameter identification of solid oxide fuel cells," International Journal of Energy Research, Wiley Online Library, vol. 45, no. 12, pp. 17903-17923, 2021,.

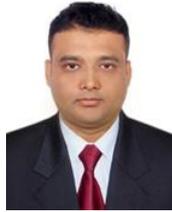

**Md. Naimur Rahman** is an Associate Professor at the Department of Electrical and Electronics Engineering of the Patuakhali Science and Technology University (PSTU), Bangladesh. He is the author of about 25 research journal articles, nearly 12 conference articles, and a book chapter on various topics related to antennas, microwaves, IoT, and electromagnetic radiation. Thus far, His google scholar citation is 213 and H-index is 8. He is now handling many research projects from the PSTU, Ministry of Science and Technology, Ministry of Education, and University Grants Commission (UGC) of Bangladesh. His research interests include IoT, antenna design, and Sensor design.

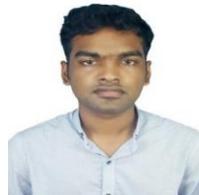

**Shafak Shahriar Sozol** worked as a research fellow and research assistant in the Institute of Information and Communication Technology, Bangladesh University of Engineering and Technology, Dhaka-1000, Bangladesh. He completed B.Sc. (Engg.) degree in Computer Science and Engineering from Patuakhali Science and Technology University, Dumki, Patuakhali-8602, Bangladesh. He is currently pursuing M.Sc. (Engg.) degree in the Institute of Information and Communication Technology at the Bangladesh University of Engineering and Technology, Dhaka-1000, Bangladesh. His research interests include machine learning, internet of things, computer vision, pattern recognition, neural networks, computer networking, distributed sensor networks, image processing, embedded system design, and data analytics. He published several IEEE conferences papers and journals.

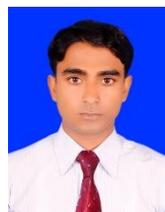

**Md. Samsuzzaman** was born in Jhenaidah Bangladesh in 1982**.** He received B.Sc. and M. Sc. Degree in Computer Science and Engineering from Islamic University Kushtia, Bangladesh in 2005 and 2007, respectively and the Ph.D. degree from the Universiti Kebangsaan Malaysia, Malaysia 2015.From February, 2008 to February, 2011, he worked as a lecturer at Patuakhali Science and Technology University (PSTU), Bangladesh. From February 2011 to August 2015. Now he is working as a post-Doctoral fellow at Universiti Kebangsaan Malaysia. He has authored or coauthored approximately 80 referred journals and conference papers. His research interests include the communication antenna design, satellite antennas, Microwave Imaging, Satellite communication, WSN and Semantic Web.

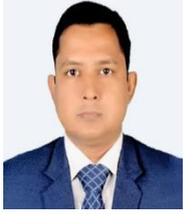

**Md. Shahin Hossin** is an Associate Professor at the Department of Soil Science of the Patuakhali Science and Technology University (PSTU), Bangladesh. He is the author of about 18 research journal articles on various topics related to soil science, soil fertility and nutrition management. He is now handling many research projects from the PSTU, Ministry of Science and Technology, and University Grants Commission (UGC) of Bangladesh. His research interests include soil fertility and nutrient management of crops in coastal soils.

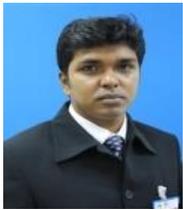

**Mohammad Tariqul Islam** is a Professor at the Department of Electrical, Electronic and Systems Engineering of the Universiti Kebangsaan Malaysia (UKM). As an expert researcher in the field of Telecommunication Engineering, he has carried out research in the areas of communication antenna design, radio astronomy antennas, satellite antennas, and electromagnetic radiation analysis. He is also the group leader of the Radio Astronomy Informatics group at UKM. He is a senior member of the IEEE, regular member of the Applied Computational Electromagnetic Society (ACES) and serving as the Editor-in-Chief of the International Journal of Electronics & Informatics (IJEI). Prof. Tariqul has been very promising as a researcher, with the achievement of several National and International Gold Medal awards, a Best Invention in Telecommunication award and a Special Award from Vietnam for his research and innovation. His publication includes over 212 research journal articles, nearly 161 conference articles, and few book chapters on various topics related to antennas, microwaves and electromagnetic radiation analysis with 7 inventory patents filed. Thus far, his publications have been cited 1717 times and his H-index is 25(Source: Scopus). In Google Scholar, the citation is 1752 and H-index is 25.

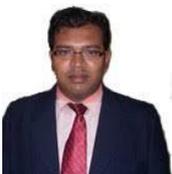

**Dr. S.M. Taohidul Islam** is working as Professor in the Electrical and Electronics Engineering Department at Patuakhali Science and Technology University, Bangladesh. He has authored or co-authored a number of referred journals and conference papers.His research interests include disaster management with information technology, geotechnicalinvestigations, geo-electric engineering, soil monitoring through electrical and electronicsengineering, IT-related environmental science, and seismic signal analysis.

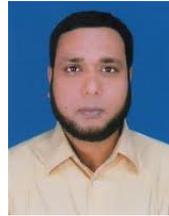

**Dr. Md. Maniruzzaman** is presently working as a Professor in Electronics and Communication Engineering (ECE) Discipline under Science, Engineering and Technology (SET) School of Khulna University (KU), Bangladesh. He received his BSc. Engg. degree in Electrical and Electronic Engineering (EEE) from Bangladesh University of Engineering and Technology (BUET), Bangladesh in 1997. And both of his M. Engg. and D. Engg. degrees are in EEE from Kitami Institute of Technology, Japan in 2003 and 2007, respectively. Dr. Maniruzzaman had been serving as the Head of ECE Discipline from January 2009 to August 2015. He also served as the Head of Statistics Discipline of Khulna University. His research area/interests include (i) Metallization for Si-ULSI Circuits/ Thin Film Technology, (ii) Noise Modeling for Interconnects in Deep Sub-Micron VLSI Circuits, (iii) Biomedical Signal/ Image Processing and Analysis, (iv) Brain Computer Interface (BCI), and (v) Internet of things (IoT), e-learning, e-health, etc.